\documentclass[prx,twocolumn,aps]{revtex4-2}
\usepackage{amsmath,bm,epsfig,color,graphicx,amsfonts,enumerate,hyperref,enumitem}
\usepackage{epsfig}
\usepackage{bm}

\usepackage[latin1]{inputenc}
\renewcommand{\=}{\!=\!}

\usepackage{xr}

\usepackage[usenames]{xcolor}
\hypersetup{
   colorlinks,
   linkcolor={blue!50!black},
   citecolor={blue!50!black},
   urlcolor={blue!80!black}
}

\begin{document}
\title{The dynamics of crack front waves in 3D material failure}
\author{Sanhita Das}
\email{Contributed equally}
\author{Yuri Lubomirsky}
\email{Contributed equally}
\author{Eran Bouchbinder}
\email{eran.bouchbinder@weizmann.ac.il}
\affiliation{Chemical and Biological Physics Department, Weizmann Institute of Science, Rehovot 7610001, Israel}

\begin{abstract}
Crack front waves (FWs) are dynamic objects that propagate along moving crack fronts in 3D materials. We study FW dynamics in the framework of a 3D phase-field framework that features a rate-dependent fracture energy $\Gamma(v)$ ($v$ is the crack propagation velocity) and intrinsic lengthscales, and quantitatively reproduces the high-speed oscillatory instability in the quasi-2D limit. We show that in-plane FWs feature a rather weak time dependence, with decay rate that increases with $d\Gamma(v)/dv\!>\!0$, and largely retain their properties upon FW-FW interactions, similarly to a related experimentally-observed solitonic behavior. Driving in-plane FWs into the nonlinear regime, we find that they propagate slower than predicted by a linear perturbation theory. Finally, by introducing small out-of-plane symmetry-breaking perturbations, coupled in- and out-of-plane FWs are excited, but the out-of-plane component decays under pure tensile loading. Yet, including a small anti-plane loading component gives rise to persistent coupled in- and out-of-plane FWs.
\end{abstract}

\maketitle

{\em Introduction}.---Material failure is a highly complex phenomenon, involving multiple scales, strong spatial localization and nonlinear dissipation. It is mediated by the propagation of cracks, which feature nearly singular stresses near their edges~\cite{freund,99bro}. In brittle materials, they reach velocities comparable to elastic wave-speeds, hence also experience strong inertial effects. In thin, quasi-2D samples, a crack is viewed as a nearly singular point that propagates in a 2D plane and leaves behind it a broken line. In thick, fully-3D samples, a crack is a nearly singular front (line) that evolves in a 3D space and leaves behind it a broken surface. While significant recent progress has been made in understanding dynamic fracture in 2D~\cite{bouchbinder.14,Chen2017,Lubomirsky2018,vasudevan2021oscillatory}, our general understanding of dynamic fracture in 3D remains incomplete~\cite{rice1985first,willis1997three,ramanathan.97,morrissey1998,morrissey2000perturbative,sharon2001propagating,sharon2002crack,fineberg2003crack,livne2005universality,ravi1998dynamic,fineberg.99,bonamy2003interaction,bonamy2005dynamic,baumberger2008magic,henry2010study,pons2010helical,henry2013fractographic,willis2013crack,adda-bedia.13,Chen2015,kolvin2015crack,bleyer2017dynamic,bleyer2017microbranching,kolvin2017nonlinear,kolvin2018topological,fekak2020crack,roch2022dynamic,steinhardt2022material,wang2022hidden,wang2023dynamics}.

A qualitative feature that distinguishes 2D from 3D material failure is the emergence of crack front waves (FWs) in the latter. FWs are compact objects that persistently propagate along crack fronts~\cite{willis1997three,ramanathan.97,morrissey1998,morrissey2000perturbative,sharon2001propagating,sharon2002crack,fineberg2003crack,livne2005universality}. In the most general case, FWs feature both a component in the main crack plane and an out-of-plane component~\cite{sharon2001propagating,sharon2002crack,fineberg2003crack}. A linear perturbation theory of singular tensile cracks, featuring no intrinsic lengthscales and rate-independent fracture-related dissipation, predicts the existence of non-dispersive in-plane FWs, whose velocity is close to the Rayleigh wave-speed $c_{_{\rm R}}$~\cite{ramanathan.97,morrissey1998}. An extended linear perturbation theory also predicts the existence of non-dispersive out-of-plane FWs in the same velocity range~\cite{adda-bedia.13}, albeit to linear order the in- and out-of-plane components are decoupled.
%%%%%%%%%%%%%% Figure %%%%%%%%%%%%%%%%%%%
\begin{figure}[ht!]
\centering
\includegraphics[width=0.97\columnwidth]{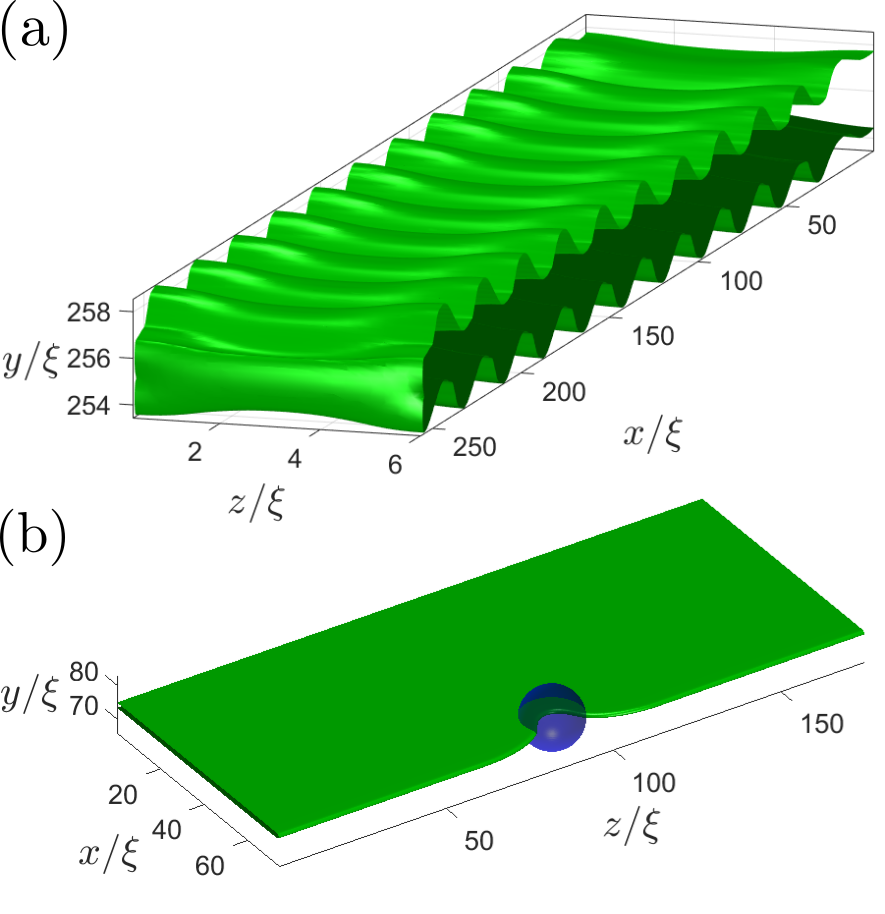}
\vspace{-0.3cm}
\caption{(a) The high-speed oscillatory instability observed in 3D phase-field simulation with $L_z\!=\!6\xi$. The crack propagates in the $x$ direction, a tensile (mode I) loading is applied in the $y$ direction and traction-free boundary conditions are employed in $z$. Plotted is the phase-field $\phi({\bm x},t)\!=\!1/2$ iso-surface. (b) A steady-state planar crack under tensile loading in a thick 3D system (with periodic boundary conditions in $z$) interacts with a tough spherical asperity (see text for details).}
\label{fig:fig1}
\end{figure}
%%%%%%%%%%%%%%%%%%%%%%%%%%%%%%%%%%%%%%%%%%%%%%%

Here, we study FWs in a 3D theoretical-computational framework that has recently quantitatively predicted the high-speed oscillatory instability in 2D~\cite{Chen2017,Lubomirsky2018,vasudevan2021oscillatory}. It is based on a phase-field approach to fracture~\cite{karma2001phase, Karma2004,Henry2004,hakim2005crack,Henry.08,Hakim.09,aranson2000continuum,eastgate2002fracture}, where large scale elastic deformations --- described by an elastic energy density $e({\bm u})$ (here ${\bf u}({\bm x},t)$ is the displacement field) --- are coupled on smaller scales near the crack edge to an auxiliary scalar field --- the phase-field $\phi({\bf x},t)$ --- that mathematically mimics material breakage. The main merit of the approach is that the dissipative dynamics of $\phi({\bm x},t)$ spontaneously generate the traction-free boundary conditions defining a crack, and consequently select its trajectory and velocity $v$. Moreover, it also incorporates intrinsic lengthscales near the crack edge --- most notably a dissipation length $\xi$ (sometimes termed the ``process zone'' size~\cite{freund,99bro}) and possibly a nonlinear elastic length $\ell_{\rm nl}$ (embodied in $e({\bm u})$~\cite{bouchbinder.14,Chen2017,Lubomirsky2018,vasudevan2021oscillatory}) --- absent in singular crack models, and a rate-dependent fracture energy $\Gamma(v)$ that accompanies the regularization of the edge singularity.

{\em The theoretical-computational framework and the quasi-2D limit}.--- We consider a homogeneous elastic material in 3D, where $L_z$ is the thickness in the $z$ direction, $L_y$ is the height in the tensile loading $y$ direction and $x$ is the crack propagation direction (we employ a treadmill procedure to obtain very long propagation distances using a finite simulation box length $L_x$~\cite{vasudevan2021oscillatory}). We use a constitutively-linear energy density $e({\bm u})\=\frac{1}{2}\lambda\,\text{tr}^2({\bm E})+ \mu\,\text{tr}({\bm E})$, with Lam\'e coefficients $\lambda$ and $\mu$ (shear modulus), and where ${\bm E}\=\tfrac{1}{2}[{\bm \nabla}{\bm u}\!+\!({\bm \nabla}{\bm u})^{\rm T}\!+\!({\bm \nabla}{\bm u})^{\rm T}{\bm \nabla}{\bm u}]$ is the Green-Lagrange metric strain tensor. The latter ensures rotational invariance, yet it introduces geometric nonlinearities (last term on the right-hand-side). However, the associated nonlinear elastic lengthscale $\ell_{\rm nl}$ remains small (unless otherwise stated~\cite{SM}), such that we essentially consider a linear elastic material and the dissipation length $\xi$ is the only relevant intrinsic lengthscale. The latter emerges once $e({\bm u})$ is coupled to the phase-field $\phi({\bm x},t)$~\cite{Chen2017,Lubomirsky2018,vasudevan2021oscillatory}.

Applying this framework in 2D, $L_z\=0$, the high-speed oscillatory instability --- upon which a straight crack loses stability in favor of an oscillatory crack when surpassing a critical velocity close to $c_{_{\rm R}}$ --- was predicted, in quantitative agreement with thin-sample experiments~\cite{Livne.07,bouchbinder.09b,Goldman2012,bouchbinder.14,Chen2017,Lubomirsky2018,vasudevan2021oscillatory}. In Fig.~\ref{fig:fig1}a, we present a high-speed oscillatory instability in a thin 3D material, $L_z\!>\!0$, where all quantities --- including the wavelength of oscillations --- agree with their 2D counterparts. These results support the validity of the 3D framework as it features the correct quasi-2D limit.

Next, we aim at exciting FWs and studying their dynamics. We consider thick systems (with $L_z/\xi\!\gg\!1$ and periodic boundary conditions along $z$), see Fig.~\ref{fig:fig1}b. Loading boundary conditions $u_i(x,y\=0,z)$ and $u_i(x,y\=L_y,z)$ are applied. In most, but not all, cases (see below), we apply tensile boundary conditions $u_y(x,y\=0,z)\=-u_y(x,y\=L_y,z)\=\delta/2$, resulting in mode I cracks initially located at the $y\=L_y/2$ plane. The tensile strain $\delta/L_y$ translates into a crack driving force $G$ (energy release rate)~\cite{freund,99bro,fineberg.99}, which is balanced by a rate-dependent fracture energy $\Gamma(v)$. The latter features $d\Gamma(v)/dv\!>\!0$, whose magnitude depends on the relxation/dissipation timescale $\tau$ of the phase-field $\phi$~\cite{vasudevan2021oscillatory}, through the dimensionless parameter $\beta\!\equiv\!\tau c_{\rm s}/\xi$ (where $c_{\rm s}$ is the shear wave-speed). The entire theoretical-computational framework depends on two dimensionless parameters, $\beta$ and $e_{\rm c}/\mu$, where $e_{\rm c}$ is the onset of dissipation energy density~\cite{vasudevan2021oscillatory}.

FWs are excited by allowing a steady-state crack front to interact with tough spherical asperities (one or more), see Fig.~\ref{fig:fig1}b. Each spherical asperity is characterized by a radius $R$ and a dimensionless fracture energy contrast $\delta\Gamma\!\equiv\!\Delta\Gamma/\Gamma_0\!>\!0$, where $\Gamma_0\!\equiv\!\Gamma(v\!\to\!0)$. The position of the asperities with respect to the crack plane, $y\=L_y/2$, determines the type of perturbation induced, i.e.~in-plane or coupled in- and out-of-plane perturbations. The resulting perturbed crack front is then described by an evolving line ${\bm f}(z,t)\=(f_x(z,t),f_y(z,t))$ parameterized by the $z$ coordinate and time $t$ (assuming no topological changes take place). Here, $f_x(z,t)$ is the in-plane component and $f_y(z,t)$ is the out-of-plane component, and an unperturbed tensile crack corresponds to ${\bm f}(z,t)\=(vt,0)$.
%%%%%%%%%%%%%% Figure %%%%%%%%%%%%%%%%%%%
\begin{figure}[ht!]
\centering
\includegraphics[width=0.97\columnwidth]{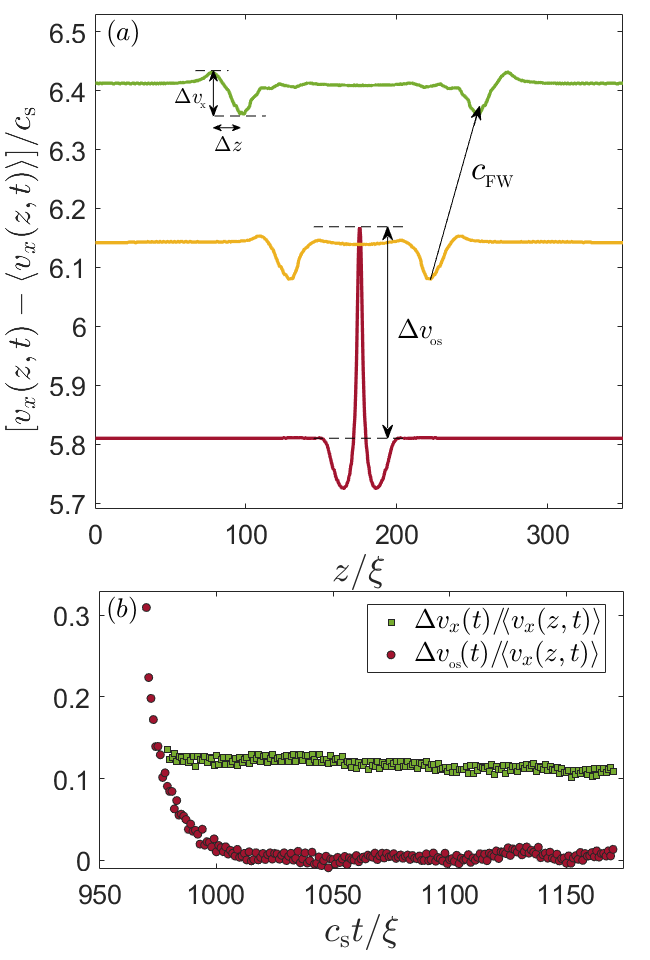}
\vspace{-0.2cm}
\caption{(a) Equal time interval snapshots of $v_x(z,t)-\langle v_x(z,t)\rangle$ (normalized and shifted for visual clarity~\cite{SM}) during in-plane FWs formation and propagation (time snapshots correspond to $t\!=\!968, 1023, 1068\,\xi/c_{\rm s}$). The velocity overshoot $\Delta{v}_{_{\rm os}}$, and FW amplitude $\Delta{v}_x$, width $\Delta{z}$ and propagation velocity $c_{_{\rm FW}}$ are all marked (see also text). FWs were generated using $v\!=\!0.6c_{\rm s}$, $R\!=\!6\xi$ and $\delta\Gamma\!=\!0.6$, and feature $c_{_{\rm FW}}\!=\!0.977c_{_{\rm R}}$. (b) $\Delta{v}_{_{\rm os}}(t)/\langle v_x(z,t)\rangle$ and $\Delta{v}_x(t)/\langle v_x(z,t)\rangle$ (see legend). See also {\bf MovieS1}-{\bf MovieS2} (\href{https://www.weizmann.ac.il/chembiophys/bouchbinder/sites/chemphys.bouchbinder/files/uploads/SupMat/front_wave_vids/Movies_SM.rar}{Download Supplementary Movies}).}
\label{fig:fig2}
\end{figure}
%%%%%%%%%%%%%%%%%%%%%%%%%%%%%%%%%%%%%%%%%%%%%%%

{\em The dynamics of in-plane FWs}.---In-plane FWs are excited by placing a single asperity whose center coincides with the crack plane, $y\=L_y/2$ (cf.~Fig.~\ref{fig:fig1}b). The tough asperity locally retards the crack front, leading to a local increase in the front curvature and $G$~\cite{rice1985first,kolvin2015crack}. The front then breaks the asperity (cf.~Fig.~\ref{fig:fig1}b), leading to a subsequent velocity overshoot $\Delta{v}_{\rm os}(t)$ ahead of the asperity (cf.~Fig.~\ref{fig:fig2}a). To quantify in-plane FWs dynamics, we employ $v_x(z,t)\!\equiv\!\partial_t f_x(z,t)$, typically with respect to $\langle v_x(z,t)\rangle\!\approx\!v$, where $\langle\cdot\rangle$ corresponds to an average along $z$ (unless otherwise stated). Strictly speaking, the physically relevant quantity is the normal front velocity, $v_{_{\!\perp}}\!(z,t)\=v_x(z,t)/\!\sqrt{1+(\partial_z f_x(z,t))^2}$. However, for our purposes here $v_x(z,t)$ itself is sufficient.

After $\Delta{v}_{\rm os}(t)$ reaches a maximum, it decays to zero (cf.~Fig.~\ref{fig:fig2}b) and a pair of in-plane FWs is generated. Each FW features an amplitude $\Delta{v}_x(t)$ (defined as the crest-to-trough difference), a width $\Delta{z}(t)$ (the corresponding crest-to-trough $z$ distance) and a propagation velocity $c_{_{\rm FW}}$ (in the laboratory frame of reference), all marked in ~Fig.~\ref{fig:fig2}a. The dimensionless FW amplitude $\Delta{v}_x(t)/\langle v_x(z,t)\rangle$ is plotted in~Fig.~\ref{fig:fig2}b. The FW inherits its scale from $R$, as shown in~\cite{SM}.

A linear perturbation theory~\cite{ramanathan.97}, developed to leading order in $|\partial_z f_x(z,t)|\!\ll\!1$, predicted the existence of non-dispersive in-plane FWs, in the absence of intrinsic lengthscales ($\xi\!\to\!0$) and for a rate-independent fracture energy ($d\Gamma(v)/dv\=0$). The theory predicts $0.94\!<\!c_{_{\rm FW}}(v)/c_{_{\rm R}}\!<\!1$ (when $v$ varies between $0$ and $c_{_{\rm R}}$). These predictions have been subsequently supported by boundary-integral method simulations of a rate-independent cohesive crack model~\cite{morrissey1998}. In~\cite{ramanathan.97}, an effective crack propagation equation of motion has been conjectured for the $d\Gamma(v)/dv\!\ne\!0$ case, suggesting that for $d\Gamma(v)/dv\!>\!0$ in-plane FWs undergo some form of attenuation during propagation.

\begin{figure}[ht!]
\centering
\includegraphics[width=0.94\columnwidth]{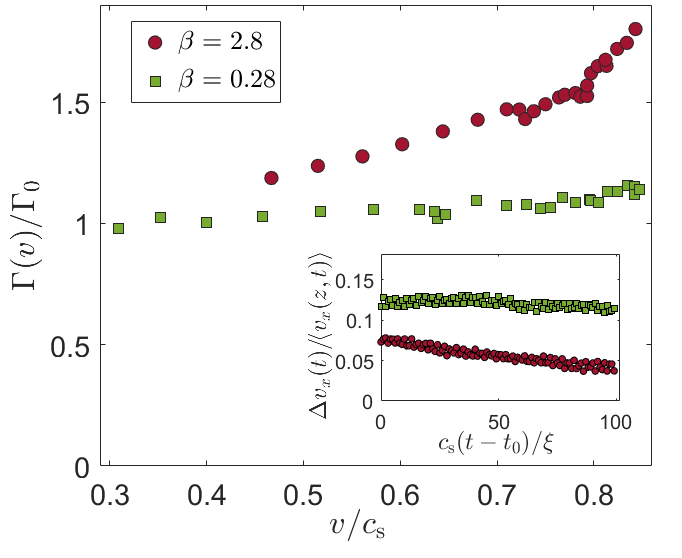}
\vspace{-0.2cm}
\caption{$\Gamma(v)/\Gamma_0$ for $\beta\!=\!0.28$ (green squares) and $\beta\!=\!2.8$ (brown circles) as previously obtained in 2D (data as in Fig.~3a in~\cite{vasudevan2021oscillatory}), where $d\Gamma(v)/dv$ differs by a factor of $4.6$. (inset) The corresponding dimensionless FW amplitude $\Delta{v}_x(t)/\langle v_x(z,t)\rangle$ for both $\beta\!=\!0.28$ (green squares) and $\beta\!=\!2.8$ (brown circles) for $v\!=\!0.6c_{\rm s}$ (FWs were generated using $R\!=\!6\xi$ and $\delta\Gamma\!=\!0.6$). In both cases, $\Delta{v}_x(t)/\langle v_x(z,t)\rangle\!\sim\!1-(t-t_0)/T$, where $1/T$ differs by a factor of $5.8$ (see text for details and discussion). $t_0$ is the time at which well-defined FWs first exist.}
\label{fig:fig3}
\end{figure}

As materials feature a rate-dependent fracture energy $\Gamma(v)$, it is important to shed light on this physical issue. Our framework naturally enables it as $d\Gamma(v)/dv$ is directly controlled by $\beta$. The evolution of the FW amplitude $\Delta{v}_x(t)/\langle v_x(z,t)\rangle$ presented in Fig.~\ref{fig:fig2} corresponds to very weak rate dependence, shown in Fig.~\ref{fig:fig3}a for $\beta\=0.28$. Such a flat $\Gamma(v)$ is characteristic of nearly ideally brittle materials such as silica glass (cf.~the experimental data in Fig.~2b of~\cite{sharon.99}). $\Delta{v}_x(t)/\langle v_x(z,t)\rangle$ in this case, presented again in the inset of Fig.~\ref{fig:fig3}, reveals a weak linear attenuation proportional to $1-(t-t_0)/T$, where $c_{\rm s}T/\xi\!\simeq\!1210$. However, while our system width $L_z$ is large enough to resolve FW propagation distances several times larger than their characteristic width $\Delta{z}$ (cf.~Fig.~\ref{fig:fig2}a), the overall propagation time $\Delta{t}$ prior to FW-FW interaction (through the periodic boundary condition, to be discussed below) is $\Delta{t}\!\sim\!{\cal O}(100)$ (cf.~Fig.~\ref{fig:fig2}b), implying $\Delta{t}\!\ll\!T$. Consequently, the presented results cannot tell apart an exponential decay from a linear one as $\exp[-\Delta{t}/T]\!\simeq\!1-\Delta{t}/T$ for $\Delta{t}\!\ll\!T$.

To address this point, and more generally the effect of the magnitude of $d\Gamma(v)/dv$ on in-plane FW dynamics, we increased $\beta$ by an order of magnitude, setting it to $\beta\=2.8$. The resulting $\Gamma(v)$, shown in Fig.~\ref{fig:fig3} (previously reported for our model in 2D~\cite{vasudevan2021oscillatory}), indeed reveals a significantly larger $d\Gamma(v)/dv$, nearly a factor 5 larger than that for $\beta\=0.28$. The emerging $d\Gamma(v)/dv$ is similar to the one observed in brittle polymers (e.g., PMMA, cf.~Fig.~2a in~\cite{sharon.99}) and in brittle elastomers (e.g., polyacrylamide, cf.~Fig.~2B in~\cite{livne2010}). The corresponding $\Delta{v}_x(t)/\langle v_x(z,t)\rangle$ is shown in the inset of Fig.~\ref{fig:fig3}, again following a linear attenuation proportional to $1-(t-t_0)/T$, this time with $c_{\rm s}T/\xi\!\simeq\!208$. Since in this case $\Delta{t}$ is comparable to $T$, the results support a linear decay, in turn implying that in-plane FWs may propagate many times their characteristic width $\Delta{z}$ even in materials with a finite $d\Gamma(v)/dv$. Moreover, we note that the decay rate $1/T$ varies between the two $\beta$ values by a factor that is comparable to the corresponding variability in $d\Gamma(v)/dv$, indeed suggesting a relation between these two physical quantities~\cite{ramanathan.97}.
%%%%%%%%%%%%%% Figure %%%%%%%%%%%%%%%%%%%%%%%%
%%%%%%%%%%%%%%%%%%%%%%%%%%%%%%%%%%%%%%%%%%%%%%%
%%%%%%%%%%%%%% Figure %%%%%%%%%%%%%%%%%%%%%%%%%
\begin{figure}[ht!]
\centering
\includegraphics[width=0.92\columnwidth]{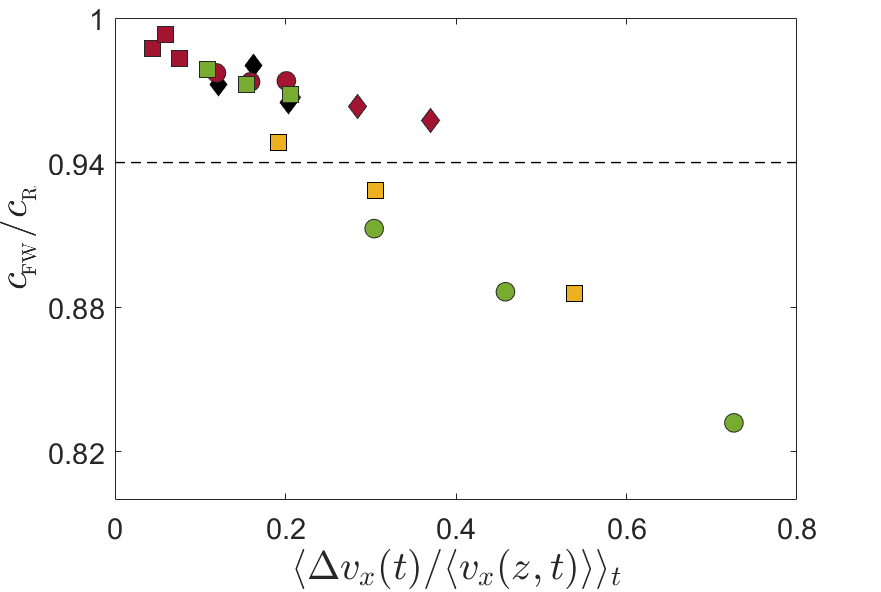}
\vspace{-0.2cm}
\caption{$c_{_{\rm FW}}/c_{_{\rm R}}$ vs.~$\langle\Delta{v}_x(t)/\langle v_x(z,t)\rangle\rangle_t$ ($\langle\cdot\rangle_t$ is a time average, prior to FW-FW interaction effects). The FW generation parameters: $v\!=\!0.5c_{\rm s}$ (diamonds), $v\!=\!0.6c_{\rm s}$ (circles), $v\!=\!0.7c_{\rm s}$ squares, $R\!=\!4\xi$ (black), $R\!=\!6\xi$ (brown), $R\!=\!12\xi$ (green), $R\!=\!18\xi$ (orange). For fixed $v$ and $R$, $\langle\Delta{v}_x(t)/\langle v_x(z,t)\rangle\rangle_t$ increases with $\delta\Gamma$~\cite{SM}.}
\label{fig:fig4}
\end{figure}
%%%%%%%%%%%%%%%%%%%%%%%%%%%%%%%%%%%%%%%%%%%%%%%

We next consider the FW velocity $c_{_{\rm FW}}$ and the possible effect of $\Delta{v}_x(t)/\langle v_x(z,t)\rangle$ on it. As explained above, the linear perturbation theory of~\cite{ramanathan.97} predicts $0.94\!<\!c_{_{\rm FW}}/c_{_{\rm R}}\!<\!1$. Consequently, we expect our excited in-plane FWs to feature $c_{_{\rm FW}}/c_{_{\rm R}}$ within this range when $\Delta{v}_x(t)/\langle v_x(z,t)\rangle$ is small. This is indeed the case in Fig.~\ref{fig:fig4}, where the dimensionless FW amplitude is controlled by systematically varying $v$, and the asperity parameters $R$ and $\delta\Gamma$ (in fact, we find that the amplitude varies linearly with $\delta\Gamma$ for fixed $v$ and $R$~\cite{SM}). However, when the amplitude is no longer small, apparently beyond the linear perturbation regime, we find that $c_{_{\rm FW}}/c_{_{\rm R}}$ decreases below 0.94, indicating that nonlinear effects tend to slow down in-plane FWs.
%%%%%%%%%%%%%% Figure %%%%%%%%%%%%%%%%%%%%%%%%%
\begin{figure}[ht!]
\centering
\includegraphics[width=0.92\columnwidth]{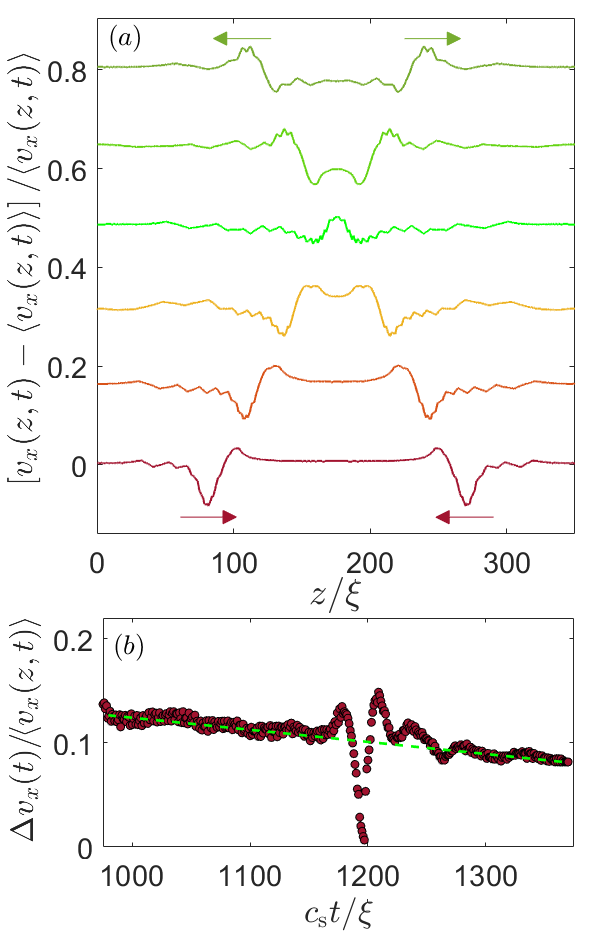}
\vspace{-0.2cm}
\caption{(a) Equal time interval snapshots (see $y$ axis label) revealing the interaction of the two FWs previously shown in Fig.~\ref{fig:fig2}a. For improved visibility, we rotate the system along the $z$ axis by $L_z/2$ such that the interaction event takes place in the middle of the system. (b) $\Delta{v}_x(t)/\langle v_x(z,t)\rangle$ for the dynamics shown in panel (a), the dashed line is a guide to the eye. See text for discussion. See also {\bf MovieS3} (\href{https://www.weizmann.ac.il/chembiophys/bouchbinder/sites/chemphys.bouchbinder/files/uploads/SupMat/front_wave_vids/Movies_SM.rar}{Download Supplementary Movies}).}
\label{fig:fig5}
\end{figure}
%%%%%%%%%%%%%%%%%%%%%%%%%%%%%%%%%%%%%%%%%%%%%%%

Finally, we take advantage of the $z$-periodic boundary conditions to study FW-FW interactions. In Fig.~\ref{fig:fig5}a, we present the interaction dynamics between the in-plane FWs previously shown in Fig.~\ref{fig:fig2}a. It is observed that the FWs retain their overall shape after the interaction, yet during the interaction they do not feature a linear superposition. This behavior is quantified in Fig.~\ref{fig:fig5}b, where $\Delta{v}_x(t)/\langle v_x(z,t)\rangle$ is plotted before, during and after FW-FW interaction (before and after the interaction it is identical for the two non-interacting FWs). In this case, it is observed that before and after the FW-FW interaction, each FW follows the very same weak linear decay previously presented in Fig.~\ref{fig:fig2}b (see superimposed dashed line) and nearly drops to zero during the interaction. This soliton-like behavior is reminiscent of similar experimental observations made in relation to coupled in- and out-of-plane FWs~\cite{sharon2001propagating,sharon2002crack,fineberg2003crack}, which are discussed next.
%%%%%%%%%%%%%% Figure %%%%%%%%%%%%%%%%%%%%%%%%%
\begin{figure}[ht!]
\centering
\includegraphics[width=0.92\columnwidth]{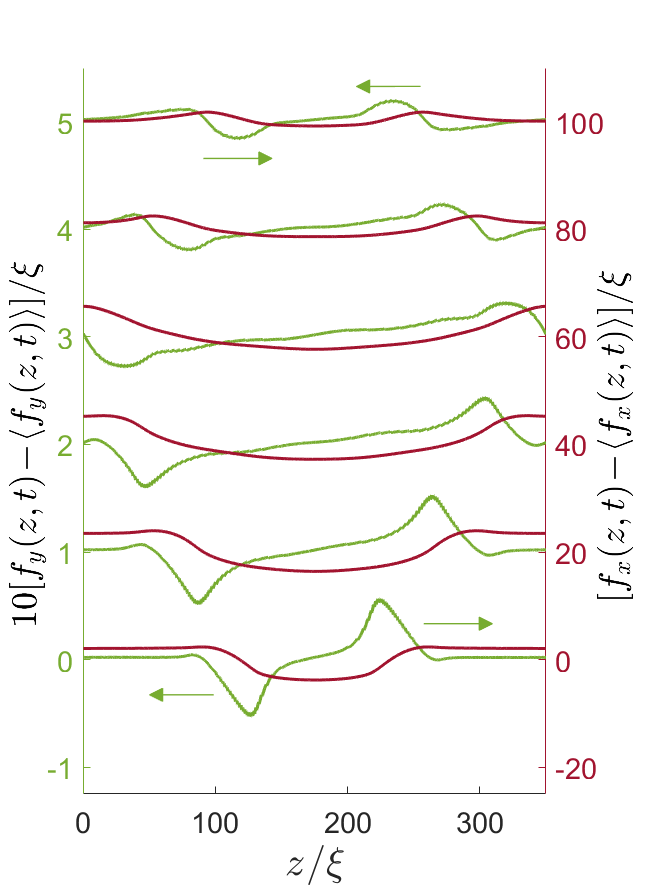}
\vspace{-0.2cm}
\caption{A pair of coupled in- and out-of-plane FWs triggered for $v\!=\!0.4c_{\rm s}$ and $\beta\!=\!0.28$ using two adjacent asperities, each characterized by $R\!=\!6\xi$ and $\delta\Gamma\!=\!0.4$. To generate an out-of-plane perturbation, one asperity is shifted by $(\delta{y}\!=\!-2\xi, \delta{z}\!=\!-2\xi)$ relative to the middle of the crack front and the other by $(\delta{y}\!=\!2\xi, \delta{z}\!=\!2\xi)$. A small anti-plane loading component is included, resulting in a mode-mixity (mode III/I) level of $3\%$ (see text for discussion). Plotted are $f_y(z,t)$ (green, multiplied by 10, see left $y$ axis) and $f_x(z,t)$ (brown, right $y$ axis) at equal time intervals. FWs persist through a FW-FW interaction, here taking place at the edges ($z\!=\!0,350\xi$) and propagate at $c_{_{\rm FW}}\!=\!0.961c_{_{\rm R}}$. See also {\bf MovieS4}-{\bf MovieS6} (\href{https://www.weizmann.ac.il/chembiophys/bouchbinder/sites/chemphys.bouchbinder/files/uploads/SupMat/front_wave_vids/Movies_SM.rar}{Download Supplementary Movies}).}
\label{fig:fig6}
\end{figure}
%%%%%%%%%%%%%%%%%%%%%%%%%%%%%%%%%%%%%%%%%%%%%%%

{\em Coupled in- and out-of-plane FWs}.---Experimentally, FWs have been observed through their fractographic signature on postmortem fracture surfaces~\cite{sharon2001propagating,sharon2002crack,fineberg2003crack,livne2005universality}, i.e.~the observed FWs featured nonlinearly coupled in- and out-of-plane components, where both $f_x(z,t)$ and $f_y(z,t)$ are non-zero and apparently propagate at the same $c_{_{\rm FW}}$. FWs in the experiments were excited by huge perturbations, 3-4 orders of magnitude larger than the out-of-plane component of the generated FWs~\cite{sharon2002crack,fineberg2003crack}, which in itself was comparable to the fracture dissipation length $\xi$. For example, asperity sizes of $100\!-\!1000\mu$m gave rise to FWs with an out-of-plane component of $0.1\mu$m in silica glass~\cite{sharon2002crack}, whose fracture dissipation (process zone) size is estimated to be in the tens of nanometers range~\cite{celarie2003glass}. Coupled in- and out-of-plane FWs are also spontaneously triggered by micro-branching events~\cite{fineberg2003crack,livne2005universality}, likely to be ``large perturbations'' as well.

Due to computational limitations --- most notably on the magnitude of $L_y$ --- we are not able to resolve this huge span in scales between the triggering perturbation and the resulting out-of-plane component. Consequently, the out-of-plane perturbations accessible to us are rather small. In particular, we perturbed the initially planar crack by a pair of adjacent asperities, one slightly shifted above the crack plane and one below, breaking the up-down symmetry. Such perturbations excite both in- and out-of-plane crack front components, but the latter decays after a short transient (while the former persists~\cite{SM}).

To understand if the latter observation is exclusively due to computational limitations (in resolving finite perturbations and the associated scale separation) or whether other physical factors are at play, we considered the recent experiments of~\cite{wang2022hidden}. It was shown therein that out-of-plane crack surface structures --- most notably surface steps~\cite{kolvin2018topological,wang2022hidden,wang2023dynamics} --- might crucially depend on the existence of small, weakly experimentally controlled, anti-plane loading component (mode III, anti-symmetric loading in the $z$ direction, e.g., due to small misalignment between the crack plane and the tensile axis). To test the possibility that a small amount of mode-mixity (mode III/I) might play a role in generating persistent coupled in- and out-of-plane FWs, we introduced a mode-mixity level of $3\%$, i.e.~$u_z(x,y\=0,z)\=-u_z(x,y\=L_y,z)\=0.03\,|u_y(x,y\=L_y,z)|$ into the above-described calculations. The results are presented in Fig.~\ref{fig:fig6}, revealing persistent propagation of a pair of coupled in- and out-of-plane FWs, featuring non-zero $f_x(z,t)$ and $f_y(z,t)$ that propagate at $c_{_{\rm FW}}\=0.961c_{_{\rm R}}$.

The amplitude of $f_y(z,t)$ is tiny, a small fraction of $\xi$ (yet it varies systematically with mode-mixity~\cite{SM}). Moreover, it is an order of magnitude small than that of $f_x(z,t)$ (notice the two $y$ axis labels in Fig.~\ref{fig:fig6}). Interestingly, this observation is consistent with experimental estimates~\cite{sharon2002crack} that suggest that $\partial_t f_y(z,t)$ is much smaller than $\partial_t f_x(z,t)$ (estimated using real-time measurements of in-plane crack velocity fluctuations at $z\=0$ and $z\=L_z$~\cite{sharon2002crack}). Overall, the observed coupled in- and out-of-plane FWs propagating at $c_{_{\rm FW}}\=0.961c_{_{\rm R}}$ with a small out-of-plane component, which also persist through FW-FW interactions, is reminiscent of several key experimental findings~\cite{sharon2001propagating,sharon2002crack,fineberg2003crack}. It remains to be seen whether a small mode-mixity, which is physically realistic, is an essential ingredient. One manifestation of it, which can be tested experimentally, is that the out-of-plane amplitude of the pair of FWs has opposite signs, see Fig.~\ref{fig:fig6}.

{\em Summary and outlook}.---Our results demonstrate that the same framework that quantitatively predicts the high-speed oscillatory instability in thin materials, also provides deep insight into FW dynamics in thick, fully 3D materials. The effect of realistic rate-dependent fracture energy $d\Gamma(v)/dv\!>\!0$ on the propagation of in-plane FWs is elucidated, as well as their solitonic nature and the effect of nonlinear amplitudes on their velocity. Persistent coupled in- and out-of-plane FWs, similar to experimental observations, are demonstrated once a small anti-plane (mode III) loading component is added to the dominant tensile (mode I) loading component.

Our findings give rise to pressing questions and subsequent investigation directions, most notably in relation to out-of-plane crack structures such as micro-branching events and surface faceting~\cite{fineberg.99,kolvin2018topological}. The roles of mode-mixity fluctuations in nominally tensile failure and of realistic material disorder/heterogeneity (we focused on homogeneous materials, discrete asperities were just introduced to generate FWs) should be particularly considered. In addition, improved computational capabilities (e.g.~based on multi-GPU implementations) should be developed in order to obtain better scale separation, which in turn may allow to understand the effect of finite out-of-plane perturbations on 3D crack dynamics.\\

{\em Acknowledgements} This work has been supported by the United States-Israel Binational Science Foundation (BSF, grant no.~2018603). E.B.~acknowledges support from the Ben May Center for Chemical Theory and Computation, and the Harold Perlman Family.

\clearpage

\onecolumngrid
\begin{center}
	\textbf{\large Supplemental Materials for:\\ ``The dynamics of crack front waves in 3D material failure''}
\end{center}

%%%%%%%%%%%%%%%%%%%%%%%%%%%%%%%%%%%%%%%%%%%%%%%%%%%%%%%%%%%%%%%%%%%%%%%%%%%%%%%%%
%%%%%%%%%%%%%%%%%%%%%% these lines of code handle the concatenation %%%%%%%%%%%%%
%%%%%%%%%%%%%%%%%%%%%%%%%%%%%%%%%%%%%%%%%%%%%%%%%%%%%%%%%%%%%%%%%%%%%%%%%%%%%%%%%
\setcounter{equation}{0}
\setcounter{figure}{0}
\setcounter{section}{0}
\setcounter{subsection}{0}
\setcounter{table}{0}
\makeatletter
\renewcommand{\theequation}{S\arabic{equation}}
\renewcommand{\thefigure}{S\arabic{figure}}
\renewcommand{\thesubsection}{S-\arabic{subsection}}
\renewcommand{\thesection}{S-\arabic{section}}
\renewcommand{\thetable}{S-\arabic{table}}
%\renewcommand*{\thepage}{S\arabic{page}}
%\renewcommand{\bibnumfmt}[1]{[S#1]}
%\renewcommand{\citenumfont}[1]{S#1}
%%%%%%%%%%%%%%%%%%%%%%%%%%%%%%%%%%%%%%%%%%%%%%%%%%%%%%%%%%%%%%%%%%%%%%%%%%%%%%%%%
%%%%%%%%%%%%%%%%%%%%%% these lines of code handle the concatenation %%%%%%%%%%%%%
%%%%%%%%%%%%%%%%%%%%%%%%%%%%%%%%%%%%%%%%%%%%%%%%%%%%%%%%%%%%%%%%%%%%%%%%%%%%%%%%%

\twocolumngrid

The goal here is to provide some technical details regarding the 3D computational framework employed in the manuscript and to offer some additional supporting data.

\subsection{The 3D phase-field model and its numerical implementation}
\label{sec:PF}

The 3D theoretical-computational framework we employed is identical to the 2D phase-field model presented in great detail in~\cite{vasudevan2021oscillatory}, extended to 3D. To the best of our knowledge, this framework is the only one that quantitatively predicted the high-speed oscillatory and tip-splitting instabilities in 2D dynamic fracture~\cite{Chen2017,Lubomirsky2018,vasudevan2021oscillatory}, and hence should serve as a basis for a 3D theory of material failure. For completeness, we briefly write down here the model's defining equations, and provide some details about the employed boundary conditions and numerical implementation in 3D.

The starting point is the Lagrangian $L\=T-U$, where the potential energy $U$ and kinetic energy $T$ are given as
\begin{eqnarray}
\label{Eq:Lagrangian_U}
U&=&\int \left[\frac{1}{2}\kappa\left(\nabla\phi\right)^{2}+ g(\phi)\,e({\bm u}) + w(\phi)\,e_{\rm c}\right]dV \ ,\\
\label{Eq:Lagrangian_T}
T&=&\int\!\frac{1}{2}f(\phi)\,\rho\left(\partial_t {\bm u}\right)^2 dV \ ,
\end{eqnarray}
in terms of the displacement vector field ${\bm u}({\bm x},t)$ and the scalar phase-field $\phi({\bm x},t)$. $dV$ is a volume differential and the integration extends over the entire system. An intact/unbroken material corresponds to $\phi\=1$, for which $g(1)\=f(1)\=1$ and $w(1)\=0$.  It describes a non-dissipative, elastic response characterized by an energy density $e({\bm u})$ on large lengthscales away from a crack edge (we use in this document `crack edge', which includes both `crack tip' in 2D and `crack front' in 3D).

The crack edge is accompanied by a large concentration of elastic energy, eventually leading to material failure, i.e.~to the loss of load-bearing capacity. This process is mathematically accounted for in the phase-field approach by the field $\phi({\bm x},t)$, which smoothly varies from $\phi\=1$ (intact/unbroken material) to $\phi\=0$ (fully broken material), and by the degradation functions $g(\phi)$, $f(\phi)$ and $w(\phi)$ that depend on it. The onset of dissipation is related to the strain energy density threshold $e_{\rm c}$ in Eq.~\eqref{Eq:Lagrangian_U}. As $\phi$ decreases from unity, $g(\phi)$ is chosen such that it decreases towards zero and $w(\phi)$ is chosen such that it increases towards unity. This process mimics the conversion of elastic strain energy into fracture energy, where the broken $\phi\=0$ phase/state becomes energetically favorable from the perspective of minimizing $U$ in Eq.~\eqref{Eq:Lagrangian_U}. Throughout this work, we operationally define the crack faces, and hence also the crack front, based on the $\phi({\bm x},t)\!=\!1/2$ iso-surface.

For $\phi\=0$, the material lost its load-bearing capacity and traction-free boundary conditions are achieved. This process is associated with a lengthscale, which emerges from the combination of the energetic penalty of developing $\phi$ gradients, as accounted for by the first contribution to $U$ in Eq.~\eqref{Eq:Lagrangian_U} that is proportional to $\kappa$, and the $\phi$-dependent elastic energy density threshold for failure $(1-w(\phi))e_{\rm c}$. Consequently, the characteristic length scale is $\xi\!\equiv\!\sqrt{\kappa/2e_{\rm c}}$, setting the size of the dissipation zone near the crack edge. The degradation functions we employed, following~\cite{vasudevan2021oscillatory}, are $f(\phi)\=g(\phi)\=\phi^4$ and $ w(\phi)\=1-\phi$. Note that the choice $f(\phi)\=g(\phi)$, where $f(\phi)$ appears in the kinetic energy of Eq.~\eqref{Eq:Lagrangian_T}, ensures that elastic wave-speeds inside the dissipation zone remain constant, as extensively discussed in~\cite{Chen2017,Lubomirsky2018,vasudevan2021oscillatory}.

To account for fracture-related dissipation, the Lagrangian $L\=T-U$ of Eqs.~\eqref{Eq:Lagrangian_U}-\eqref{Eq:Lagrangian_T} is supplemented with the following dissipation function (directly related to the phase-field $\phi({\bm x},t)$)
\begin{equation}
D \equiv \frac{1}{2\chi}\int \left(\partial_t \phi\right)^{2}dV \ ,
\label{Eq:dissipation}
\end{equation}
where $\chi$ is a dissipation rate coefficient that determines the rate-dependence of the fracture energy $\Gamma(v)$. The quasi-static fracture energy, $\Gamma_0\=\Gamma(v\!\to\!0)$, is proportional to $e_{\rm c}\xi$~\cite{vasudevan2021oscillatory}. The evolution of $\phi({\bm x},t)$ and ${\bm u}({\bm x},t)$ is derived from Lagrange's equations
\begin{eqnarray}
\frac{\partial}{\partial t}\left[\frac{\delta L}{\delta\left(\partial\psi/\partial t\right)}\right]-\frac{\delta L}{\delta\psi}
+\frac{\delta D}{\delta\left(\partial\psi/\partial t\right)}=0 \ ,
\label{Eq:Lagrange_eqs}
\end{eqnarray}
where $\psi\=(\phi,u_x,u_y,u_z)$, i.e.~${\bm u}\=(u_x,u_y,u_z)$ are the components of the displacement vector field.

As explained in the manuscript, we employed the following constitutively-linear elastic energy density
\begin{equation}
e({\bm u})=\frac{1}{2}\lambda\,\text{tr}^2({\bm E}) + \mu\,\text{tr}({\bm E}) \ ,
\label{eq:SVK}
\end{equation}
where ${\bm E}\=\tfrac{1}{2}[{\bm \nabla}{\bm u}+({\bm \nabla}{\bm u})^{\rm T}+({\bm \nabla}{\bm u})^{\rm T}{\bm \nabla}{\bm u}]$ is the Green-Lagrange metric strain tensor, and $\lambda$ and $\mu$ (shear modulus) are the Lam\'e coefficients. We set $\lambda\=2\mu$ in all of our calculations. Using Eqs.~\eqref{Eq:Lagrangian_U}-\eqref{Eq:dissipation}, with Eq.~\eqref{eq:SVK}, inside Eq.~\eqref{Eq:Lagrange_eqs} fully defines our field equations in 3D (that should be solved in a given 3D domain, and supplemented with proper initial and boundary conditions, as described below). The resulting equations are nondimensionalized by expressing length in units of $\xi$, time in units of $\xi/c_{\rm s}$, energy density in units of $\mu$ and the mass density $\rho$ in units of $\mu/c_{\rm s}^2$ ($c_{\rm s}$ is the shear wave-speed). Once done, the dimensionless set of equations depends on two dimensionless parameters: $e_{\rm c}/\mu$ (the ratio between the dissipation onset threshold $e_{\rm c}$ and a characteristic elastic modulus) and on $\beta\=\tau\,c_{\rm s}/\xi$ (where we defined $\tau\!\equiv\!(2\chi e_{\rm c})^{-1}$), which controls the $v$-dependence of the fracture energy, $\Gamma(v)$, as discussed in the manuscript.

As discussed extensively in~\cite{Chen2017,Lubomirsky2018,vasudevan2021oscillatory}, near crack edge elastic nonlinearity --- embodied in Eq.~\eqref{eq:SVK} in the Green-Lagrange strain tensor ${\bm E}$ --- gives rise to a nonlinear elastic lengthscale $\ell_{\rm nl}$ that scales as $\ell_{\rm nl}/\xi\!\sim\!e_{\rm c}/\mu$. In the calculations in the context of the high-speed oscillatory instability, cf.~Fig.~1a in the manuscript, we set $e_{\rm c}/\mu\=0.02$. The latter leads to a sizable nonlinear elastic lengthscale $\ell_{\rm nl}$ in the ultra-high crack propagation velocities regime considered therein ($v\!\to\!c_{_{\rm R}}$), which controls the wavelength of oscillations (note, though, that it was shown~\cite{vasudevan2021oscillatory} that the high-speed oscillatory instability persists also in the limit $\ell_{\rm nl}/\xi\!\to\!0$, where the wavelength is controlled by $\xi$). In the rest of our calculations, where the dynamics of crack front waves (FWs) were of interest, we focused on a linear elastic behavior, where $\ell_{\rm nl}$ is negligibly small. The latter is ensured by setting $e_{\rm c}/\mu\=0.005$ and considering $v\!\le\!0.7c_{\rm s}$. Consequently, as stated in the manuscript, in all of our FW-related calculations, the material is essentially linear elastic and the only relevant intrinsic lengthscale is the dissipation length $\xi$. The rate of dissipation parameter $\beta$ was varied between $\beta\=0.28$ and $\beta\=2.8$, as discussed in the manuscript.

Our calculations were performed in boxes of length $L_x$ in the crack propagation direction $x$, height $L_y$ in the loading direction $y$ and $L_z$ in the thickness direction $z$. In all of our calculations, we set $L_x\!=\!150\xi$. However, we employed a treadmill procedure (as explained in~\cite{vasudevan2021oscillatory}), which allows to simulate very large crack propagation distances. Consequently, our system is effectively infinite in the crack propagation direction. In Fig.~2a in the manuscript, where our focus was on testing the reproducibility of the high-speed oscillatory instability in the thin, quasi-2D limit, we used $L_z\=6\xi$ and a large $L_y$. This calculation also employed traction-free boundary conditions at $z\=0$ and $z\=L_z$. In the rest of our calculations, which focused on FW dynamics, we were interested in thick systems. To that aim, we used $L_z\=350\xi$ (note that in the illustrative Fig.~1b in the manuscript, we showed a smaller $L_z$ for visual clarity) and periodic boundary conditions in $z$. Due to the enormous computational cost involved in our large-scale calculations, employing such a large $L_z$ implies that $L_y$ is rather constrained. In all of the FW calculations we used $L_y\=150\xi$. The loading conditions at $y\=0$ and $y\=L_y$ are discussed in the manuscript. Note that the crack propagation velocity $v$ is set by controlling the crack driving force $G$ (through the loading conditions), following energy balance $\Gamma(v)\=G$.

The resulting field equations corresponding to Eqs.~\eqref{Eq:Lagrange_eqs}, cf.~Eqs.~(A.1)-(A.3) in~\cite{vasudevan2021oscillatory}, are spatially discretized in 3D on a cubic grid with a discretization size $\Delta{x}\=\Delta{y}\=\Delta{z}\=0.25\xi$, following the same spatial discretization scheme described in~\cite{vasudevan2021oscillatory}, straightforwardly extended from 2D to 3D. The temporal discretization (at any spatial grid point) involves different schemes for the scalar phase-field $\phi$ and the vectorial displacement field ${\bm u}$. For the former, we employ a simple forward Euler scheme $\phi_{n+1}\=\phi_{n}+\dot{\phi}_{n}\Delta{t}$ as in~\cite{vasudevan2021oscillatory}, where the subscript $n$ refers to the current time step, $t_n\=n\Delta{t}$, with $\Delta{t}$ being the discrete time step size.

For ${\bm u}$, we developed a specifically-adapted Velocity Verlet scheme. As in the conventional Velocity Verlet scheme~\cite{verlet1967computer}, the displacement ${\bm u}_{n+1}$ is given to second order in $\Delta{t}$ as ${\bm u}_{n+1}\={\bm u}_{n}+{\bm v}_{n}\Delta{t}+\tfrac{1}{2}{\bm a}_{n}\Delta{t}^2$, in terms of ${\bm u}_n$, the velocity ${\bm v}_n$ and the acceleration ${\bm a}_n$. The appearance of the degradation function $f(\phi)$ in the kinetic energy in Eq.~\eqref{Eq:Lagrangian_T} implies that ${\bm a}_{n+1}$ depends on ${\bm v}_{n+1}$ itself (cf.~Eq.~(A.3) in~\cite{vasudevan2021oscillatory}), and hence the conventional Velocity Verlet~\cite{verlet1967computer} expression for ${\bm v}_{n+1}$, i.e.~${\bm v}_{n+1}\={\bm v}_n+\tfrac{1}{2}({\bm a}_n+{\bm a}_{n+1})\Delta{t}$, cannot be used (since, as explained, ${\bm a}_{n+1}$ depends on ${\bm v}_{n+1}$). Instead, we defined an auxiliary acceleration $\tilde{\bm a}_{n+1}$ that was estimated using an auxiliary velocity $\tilde{\bm v}_{n+1}\={\bm v}_n+{\bm a}_n\Delta{t}$, from which we estimated ${\bm v}_{n+1}$ according to ${\bm v}_{n+1}\={\bm v}_n+\tfrac{1}{2}({\bm a}_n+\tilde{\bm a}_{n+1})\Delta{t}$.

This specifically-adapted Velocity Verlet scheme involved the estimation of the auxiliary acceleration $\tilde{\bm a}_{n+1}$, which entails the computation of the divergence of the stress tensor (cf.~Eq.~(A.3) in~\cite{vasudevan2021oscillatory}). The latter, whose computation is a serious bottleneck, was reused to evaluate ${\bm a}_{n+1}$ at the next time step. This reuse of the divergence of the stress gives rise to more than a two-fold speedup in run-times compared to the temporal discretization scheme used in~\cite{vasudevan2021oscillatory}, which is essential for the very demanding 3D computations. Finally, the time step size $\Delta{t}$ is set according to the $\beta$ parameter, taking into account the associated stability condition of the diffusion-like $\dot{\phi}$ equation ($\Delta{t}$ of course also satisfies the CFL condition, which is less stringent in our case).

All of our calculations are perform on a single GPU (NVIDIA TeslaV100\_SXM2, QuadroRTX8000 or QuadroRTX6000) available on WEXAC (Weizmann EXAscale Cluster), which is a large-scale supercomputing resource at Weizmann Institute of Science. Our computations are very demanding in terms of memory, typically involving $\sim\!40$GB of memory per simulation. Consequently, all data analysis has to be performed on the fly, as it is simply not practical to save snapshots of the fields. To that end, we used Matlab's C++ engine that enables to execute Matlab scripts during run-time. In order to maximize performance, our computational platform is entirely implemented using C/C++ and CUDA, with typical simulation times of a few days per simulation, depending on the parameters.

\vspace{0.5cm}
{\bf A}. {\em FWs generation and discrete heterogeneities/asperities}\\
%\label{sec:asperity}

As explained in the manuscript, FWs generation involves 3 parameters, the steady-state crack front velocity $v$, the asperity radius $R$ and its dimensionless fracture energy contrast $\delta\Gamma\!\equiv\!\Delta\Gamma/\Gamma_0$. To obtain a steadily propagating crack, we first introduced a planar crack and iteratively relaxed the elastic fields until reaching a mechanical equilibrium state under a prescribed loading. The latter corresponds to a given crack driving force $G$. Then, the crack was allowed to propagate until reaching a steady-state according to energy balance $\Gamma(v)\=G$, as explained above.

FWs are excited by allowing the steadily propagating planar crack to interact with discrete heterogeneities in the form of tough spherical asperities. To generate asperities, we introduce an auxiliary static (quenched) ``noise field'' $\zeta({\bm x})$, which can be coupled to any physical parameter in the fracture problem. This coupling is achieved by transforming an originally spatially uniform parameter $\alpha_0$ into a field of the form $\alpha({\bm x})\=\alpha_0[1+\alpha_{_{\zeta}} \zeta({\bm x})]$, where $\alpha_{_{\zeta}} $ is a coupling coefficient.

We applied this formulation to the fracture energy, whose quasi-static value scales as $\Gamma_0\!\sim\!e_{\rm c}\xi\!\sim\!\sqrt{\kappa e_{\rm c}}$, by simultaneously coupling $\kappa$, $e_{\rm c}$ and $\chi$ to $\zeta({\bm x})$, while keeping $\xi\!\sim\!\sqrt{\kappa e_{\rm c}}$ and $\tau\!\sim\!(\chi e_{\rm c})^{-1}$ fixed. This choice ensures that $\beta\=\tau c_{\rm s}/\xi\!\sim\!(\chi e_{\rm c} \xi)^{-1}$ remains fixed, i.e.~the asperities feature an overall dimensionless fracture energy contrast $\delta\Gamma\!\equiv\!\Delta\Gamma/\Gamma_0$ (controlled by $\kappa e_{\rm c}$) compared to the homogeneous surrounding material, but the very same fracture rate dependence $d\Gamma(v)/dv$ (controlled by $\beta$).

Finally, discrete spherical asperities are obtained by choosing $\zeta({\bm x})$ with a compact support in the form $\zeta({\bm x})\=(1-|{\bm x}-{\bm x}_0|/R)^5$ for $|{\bm x}-{\bm x}_0|\!\le\!R$ and $\zeta({\bm x})\=0$ elsewhere. Here ${\bm x}_0$ is the location of the center of the asperity and $R$ is its radius, as defined in the manuscript. Asperities are allowed to overlap by simply summing the contributions of the individual asperities to the noise field.

\subsection{Additional supporting results}
\label{sec:results}

\begin{figure}[ht!]
\centering
\includegraphics[width=1\columnwidth]{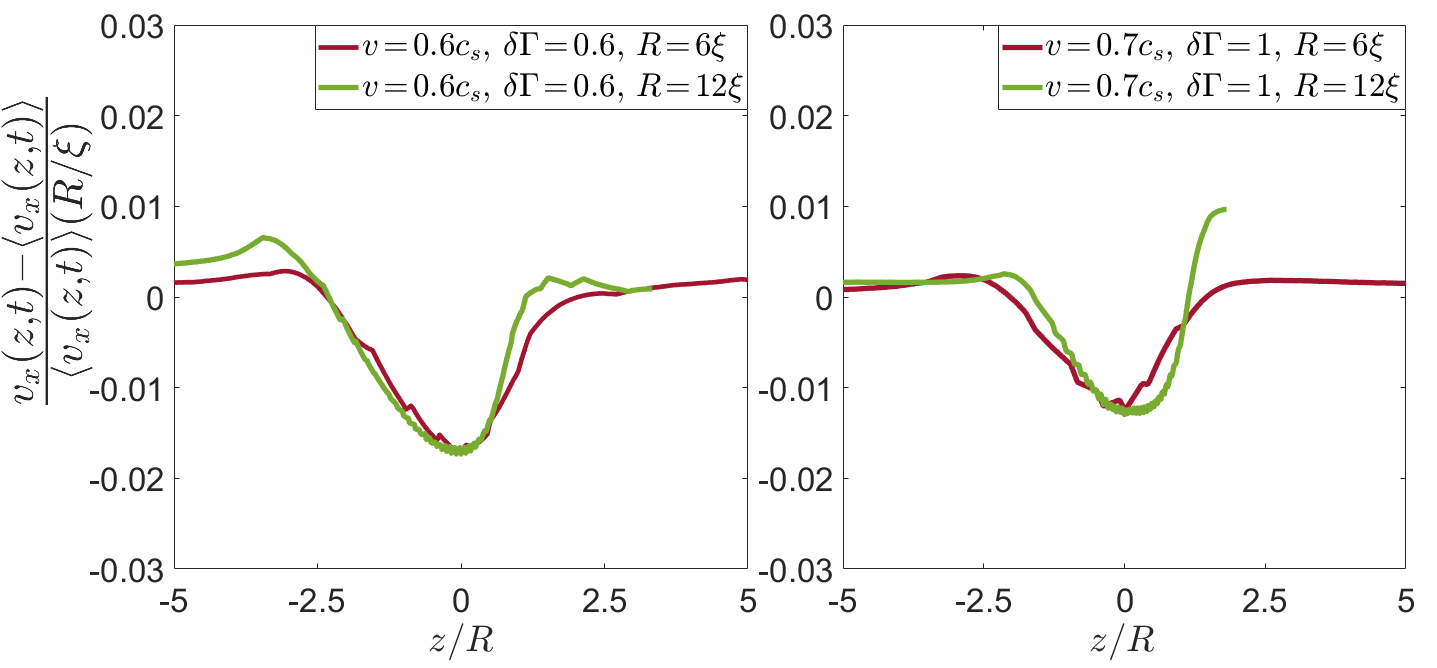}
\vspace{-0.2cm}
\caption{Two examples of the approximate scaling of both the FW dimensionless amplitude and its width with the asperity size $R$. It is demonstrated here for two cases (see legends for the FW generation parameters) by an approximate collapse obtained once the FW dimensionless amplitude is rescaled by $R/\xi$ (see $y$ axis label, which is the same for both the left and right panels) and $z$ by $R$, when $R$ is doubled (see legends). The FWs are shifted such that they are centered near $z\!=\!0$.}
\label{fig:figS1}
\end{figure}

In this section, we provide additional supporting results that are referred to in the manuscript. First, in Fig.~\ref{fig:figS1} we show that in-plane FWs approximately inherit their scale, both amplitude and width, from the asperity size $R$. This is similar to experimental findings reported in relation to the out-of-plane component of FWs~\cite{sharon2001propagating,sharon2002crack,fineberg2003crack}.
%%%%%%%%%%%%%% Figure %%%%%%%%%%%%%%%%%%%%%%%%%
%%%%%%%%%%%%%%%%%%%%%%%%%%%%%%%%%%%%%%%%%%%%%%%

\begin{figure}[ht!]
\centering
\includegraphics[width=1.1\columnwidth]{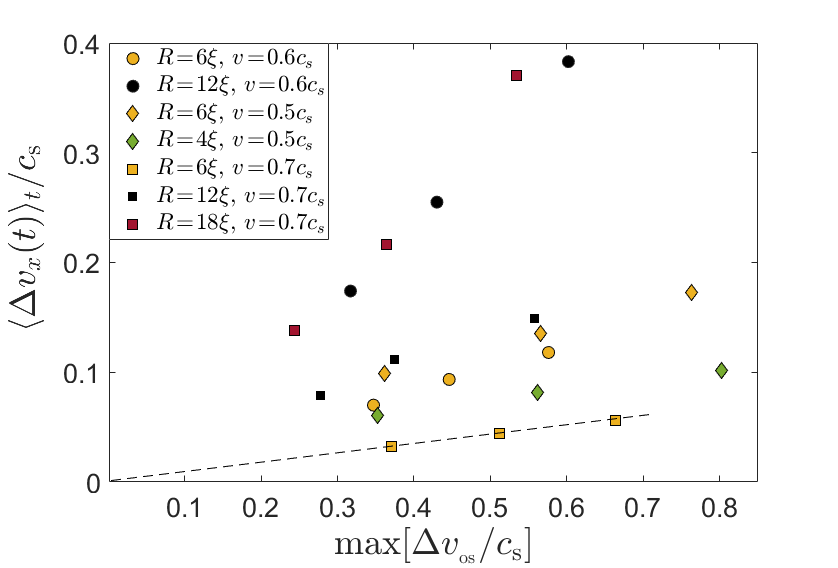}
\vspace{-0.2cm}
\caption{$\langle\Delta{v}_x\rangle_t$ (the time-averaged $\Delta{v}_x(t)$) vs.~max$\left[\Delta{v}_{_{\rm os}}\right]$ (the maximum of $\Delta{v}_{_{\rm os}}(t)$), both in units of $c_{\rm s}$, for various $R$ and $v$ values (as indicated in the legend), and variable $\delta\Gamma\!=\!0.6, 0.8, 1$. Increasing $\delta\Gamma$ linearly increases max$\left[\Delta{v}_{_{\rm os}}\right]$. The same color means the same $R$ and the same symbol means the same $v$ (see legend). The dashed line is added as a guide to the eye.}
\label{fig:figS2}
\end{figure}

In Fig.~2 in the manuscript, we showed that FW generation is accompanied by an initial velocity overshoot $\Delta{v}_{_{\rm os}}(t)$ that develops ahead of the asperity, after the latter is broken. We found that the maximal velocity overshoot, max$\left[\Delta{v}_{_{\rm os}}\right]$, controls the amplitude $\Delta{v}_x$ of the generated FW. We also found that $\Delta{v}_{_{\rm os}}$ varies approximately linearly with $\delta\Gamma$ for fixed $v$ and $R$ (not shown). In Fig.~\ref{fig:figS2}, we show that $\Delta{v}_x$ varies predominantly linearly with max$\left[\Delta{v}_{_{\rm os}}\right]$, when the latter is varied by varying $\delta\Gamma$ for fixed $v$ and $R$.
%%%%%%%%%%%%%% Figure %%%%%%%%%%%%%%%%%%%%%%%%%
%%%%%%%%%%%%%%%%%%%%%%%%%%%%%%%%%%%%%%%%%%%%%%%

\subsection{Supporting movies}
\label{sec:movies}

A major merit of the employed 3D computational framework is that it enables tracking crack evolution in 3D in real (computer) time. Consequently, we supplement the results presented in the manuscript with movies of the corresponding 3D dynamics. The Supplemental Materials include 6 movies, which can be downloaded from this link: \href{https://www.weizmann.ac.il/chembiophys/bouchbinder/sites/chemphys.bouchbinder/files/uploads/SupMat/front_wave_vids/Movies_SM.rar}{Download Supplementary Movies}, described as follows:
\begin{itemize}
\item {\bf MovieS1}: A movie that shows FW generation and propagation prior to FW-FW interaction, following Fig.~2a in the manuscript. In the latter, equal time interval snapshots were presented. The snapshots therein were shifted according to $0.006\,c_{\rm s}t/\xi$ to demonstrate FW propagation.
\item {\bf MovieS2}: The same calculation as in MovieS1 and Fig.~2 in the manuscript, here showing the phase-field $\phi({\bm x},t)\!=\!1/2$ iso-surface. Note the different scales of the axes.
\item {\bf MovieS3}: A movie that corresponds to the FW-FW interaction shown in Fig.~5a in the manuscript. In the latter, equal time interval snapshots were presented. The snapshots therein were shifted according to $0.004c_{\rm s}(t-t_0)/\xi$ to demonstrate FW propagation.
    \vspace{0.2cm}
\item {\bf MovieS4}: A movie that corresponds the coupled in- and out-of-plane perturbation induced by two asperities as in Fig.~6 in the manuscript, albeit under pure mode I (no mode III). The movie shows that coupled in- and out-of-plane components are generated by the perturbation, but that the out-of-plane component decays, while the in-plane persistently propagates.
\item {\bf MovieS5}: A movie that corresponds to Fig.~6 in the manuscript, i.e.~it is identical to MovieS4, but with a mode-mixity (mode III/I) of $3\%$. Note that in Fig.~6 in the manuscript, snapshots corresponding to the left $y$ axis were shifted according to $0.05\!\times\!0.4\,c_{\rm s}(t\!-\!t_0)/\xi$, while those corresponding to the right $y$ axis were shifted according to $0.4\,c_{\rm s}(t\!-\!t_0)/\xi$.
\item {\bf MovieS6}: The same as MovieS5, but with a mode-mixity (mode III/I) of $5\%$. The resulting coupled in- and out-of-plane FW features an out-of-plane component that approximately scales with the level of mode-mixity.
\end{itemize}

\end{document}